\documentclass[twoside]{LCWS11}
\usepackage[latin1]{inputenc}
\usepackage[dvips]{graphicx,epsfig,color}
\usepackage{wrapfig,rotating}
\usepackage{amssymb,amsmath,array}
\usepackage{cite}
\begin{document}
\title{
BDS tuning and Luminosity Monitoring in CLIC } 
\author{B. Dalena$^{1,}$$^2$, J. Barranco$^2$, A. Latina$^2$, E. Marin$^2$,\\
J. Pfingstner$^2$, D. Schulte$^2$, J. Snuverink$^2$, R. Tom\'as$^2$, G.
Zamudio$^2$ 
\vspace{.3cm}\\
1- CEA/SACLAY, DSM/Irfu/SACM - 91191 Gif-sur-Yvette - France \\ 
2- CERN, Geneva - Switzerland\\
}

\maketitle

\begin{abstract}
The emittance preservation in the Beam Delivery System (BDS) is one of 
the major challenges in CLIC. The fast detuning of the final focus 
optics requires an on-line tuning procedure in order to keep luminosity
close to the maximum.
Different tuning techniques have been applied to the CLIC BDS 
and in particular to the Final Focus System (FFS) in order to mitigate
static and dynamic imperfections.
Some of them require a fast luminosity measurement.   
Here we study the possibility to use beam-beam backgrounds processes at 
CLIC 3 TeV CM energy as fast luminosity signal. In particular the 
hadrons multiplicity in the detector region is investigated. 
\end{abstract}

\section{Introduction}

Conventional beam-based alignment techniques partially succeeded to 
tune the static imperfections in the CLIC BDS. In particular 
they have been proven successful in the collimation section 
alone while they recover only few percent of luminosity when applied 
to the CLIC FFS. 
This is due to its strong non-linear beam dynamics~\cite{Andre08} and 
very low $\beta$ function at the IP ($\beta^{*}$)~\cite{Roge09}.    

Integrated simulations of Main Linac (ML) and BDS
including ground motion lead to a luminosity loss of the order of
10$\%$ after about 1 hour~\cite{Jurg11}, according to the 
ground motion model used. The luminosity loss can be fully recovered by
scanning precomputed orthogonal tuning knobs. These tuning knobs 
consist of linear combinations of five FFS sextupole displacements 
built to control the main linear aberrations of the beam at the 
Interaction Point (IP). 
The source of the luminosity loss is therefore due to FFS detuning.  
A fast on-line tuning procedure is required in order to reduce the 
luminosity loss during operation as well as for the tune-up of the 
machine.
As we will see in section~\ref{tune},
the most successful tuning techniques exploit the luminosity as figure
of merit. 
It is mandatory to have a method to estimate luminosity variations which
can be used for machine optimization.  

The measurement of luminosity in $e^{+}e^{-}$ colliders is usually done 
by detecting radiative Bhabhas ($e^+ e^- \to e^+ e^- \gamma$)
~\cite{radBha} in the detector's forward region. In CLIC at 3 TeV CM
energy the radiative Bhabhas signal cannot be easily distinguished in
the spent beams low energy tails.\\
The low angles Bhabhas have a lower event rate than radiative Bhabhas
at the CLIC CM energy. These methods need from 7 to 70 minutes in order
to reach 1$\%$ precision in the measurement of the luminosity~\cite{
lumimeas}. 
The fast detuning of the machine is then not compatible with this 
technique.

The possibility to use secondary particles emitted during the 
beam-beam interaction 
to monitor luminosity at CLIC has already been proposed~\cite{Dani04}. 
In particular, the possibility to use the beamstrahlung photons as a 
fast luminosity signal has been exploited in~\cite{Elias06}. The 
measurement of the beam sizes at the IP, using incoherent pairs both 
alone or in combination with 
beamstrahlung, has been explored in~\cite{Yoko95}.
In the following the required luminosity measurements and the results
in terms of CLIC BDS performances are presented, for three 
different techniques studied. The beam-beam background processes and
their correlations with luminosity are studied, considering several 
beam aberrations at the IP. 
Finally, a new potential signal from the $\gamma \gamma \to$ hadrons
process is assessed for tuning purposes.

\section{BDS tuning}\label{tune}

We discuss here the results of different techniques applied to the 
CLIC BDS in order to mitigate static imperfections. We consider magnets
displacements in the horizontal and vertical plane, magnets strength and roll
errors are foreseen to be studied in the future. The main reason to study
the impact of magnets displacements only on the machine performance is because
of their relevance in the dynamic case. Detailed studies  
of dynamic imperfections can be found in~\cite{Joch11}. 

\begin{wrapfigure}{r}{0.5\columnwidth}
\centerline{\includegraphics[width=0.5\columnwidth]
{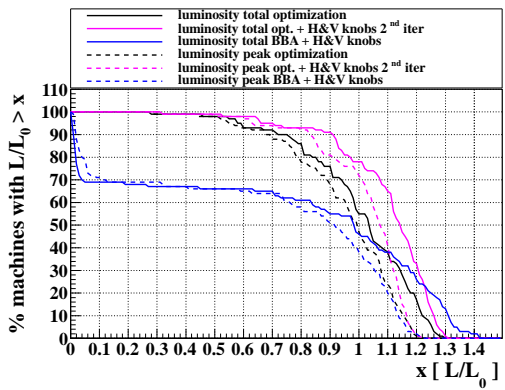}}
\caption{Luminosity distribution of 100 machines after three different
  alignment procedures starting from an initial random pre-alignment
  of 10 $\mu$m.}\label{Fig:1}
\end{wrapfigure}
The results of the tuning of the BDS against magnets displacements
are summarized in Fig.~\ref{Fig:1}. 

A random Gaussian displacement of each magnet with a $\sigma$ of 
10~$\mu$m in the two transverse planes is considered, which is
the pre-alignment specification in all the BDS.  
The number of machines reaching the target luminosity
is quite different depending on the applied techniques. 
In order to accommodate for the static and dynamic imperfections the 
CLIC BDS lattice (with L$^{*}$=3.5 m) is designed to reach a peak and 
total luminosity higher than the nominal values, by $\sim$ 20$\%$ and 
$\sim$ 30$\%$, respectively.
The target luminosity after the correction for the static imperfections
is 110$\%$, the remaining 10$\%$, to reach the design peak luminosity 
of 120$\%$, is the budget for the dynamic imperfections. 
Here three different procedures are studied: BBA in combination with tuning
knobs, luminosity optimization and luminosity optimization in combination
with tuning knobs. 

The Beam Based Alignment (BBA) technique consists of the 1-to-1 
correction followed by Dispersion Free Steering (DFS)~\cite{Raub91} 
in the vertical plane and target DFS in the horizontal one. 
In the 1-to-1 correction the beam is steered through the center of the
BPMs. DFS is a technique that measures the dispersion along the line,
using off-energy test beams, and corrects it to zero or to the nominal
value. The energy difference of 0.1$\%$ is used to measure dispersion. 
The assumed BPM resolution in these simulations is 10~nm.
This technique has proven successful in the CLIC
collimation section alone, while it fails when the FFS is also 
considered. The possibility to use tuning knobs based on linear 
combinations of sextupoles displacements has been already explored in 
CLIC. 
Knobs to control the offsets and angles at the IP, the waist shift 
and the dispersions functions were partially successful.  
New tuning knobs are built here using FFS sextupoles displacements, 
in order to control mainly couplings, dispersions and waist-shift in 
the two transverse planes. These tuning knobs, applied after BBA, 
manage to shrink the transverse beam sizes 
close to the nominal values, recovering up to 50$\%$ of luminosity 
loss in half of the different seeds used in the simulation. 
Iterations of BBA and tuning knobs improve the correction. The final 
total and peak luminosity obtained after fifth iterations of this 
technique for 100 random misaligned machines are shown in 
Fig.~\ref{Fig:1} (blue line). 
About 30$\%$ of machines reach 110$\%$ of CLIC nominal luminosity. 
Of these about 15$\%$ exceed the design value of 130$\%$ for the total 
luminosity, while this is not the case for the peak luminosity. 
This effect is explained by the smaller horizontal beam size, reached 
after the BBA and FFS knobs scan, with respect to the nominal value, 
which causes on one hand the enhancement of total luminosity and on the 
other hand the emission of more beamstrahlung photons with the consequent
increase of average energy loss that smears the luminosity spectrum in 
the energy peak.   

In the luminosity optimization procedure, all the elements of the FFS
are moved in order to maximize luminosity, using the Nedler-Mead 
algorithm (Simplex).
In this case more than 60$\%$ of the machines reach 110$\%$ of CLIC 
total nominal luminosity.
It is worth noticing that when the luminosity optimization is combined
with tuning knobs about 90$\%$ of the machines reach 90$\%$ of CLIC 
nominal total luminosity (Fig.~\ref{Fig:1}). 
The number of luminosity measurements needed by the luminosity 
optimization procedure is one order of magnitude larger than the one 
required by the BBA and Knobs scan technique.
It is therefore crucial for CLIC to be able to measure luminosity as 
fast as possible (in the order of seconds) and to be able to tune the
system in the most efficient way. 
The use of more sophisticated optimization algorithms and non linear 
knobs could improve the overall luminosity results and reduce the 
number of luminosity measurements required.\\ 
In the following we concentrate on the definition of fast luminosity 
signals. For this purpose the beam-beam background processes 
and their correlation with the main sources of luminosity degradation
are presented.

\section{Luminosity signals and colliding beam parameters} 

We study the variation of different signals 
from beam-beam interaction, according to 10 different beam
aberrations at the IP. 
The size of the aberrations is chosen to produce a luminosity loss of 
about 30$\%$. The six Signals (S) we define are: 
\begin{enumerate}
\item \texttt{coherent} -- number of coherent pairs from the two beams;
\item \texttt{(n$_{\gamma1}$+n$_{\gamma2}$)/2} -- average number of 
      beamstrahlung photons from the two beams;
\item \texttt{1.0 - $\left| \Delta \textrm{n}_{\gamma}/\Sigma 
  \textrm{n}_{\gamma} \right|$} -- difference of the number of beamstrahlung
      photons from the two beams normalized to their sum; 
\item \texttt{n$_{\gamma1}$/n$_{\gamma2}$} -- ratio of the number of beamstrahlung
      photons from the two beams; 
\item \texttt{hadrons} -- total number of $\gamma \gamma \to$ hadrons
      events;
\item \texttt{incoherent} -- total number of incoherent pairs from the two 
      beams.
\end{enumerate}
The full transport of the two beams through the main LINAC and the BDS
is simulated with the tracking code PLACET~\cite{plac}. The sextupoles of 
one beamline are displaced according to the linear knobs introduced 
in section~\ref{tune}, 
generating the beam phase space distortion at the IP. The second 
beamline instead is kept without any errors.
Five of the knobs are built by horizontal sextupoles 
displacements in order to control horizontal dispersion (D$_{x}$), 
horizontal and vertical waist-shift ($\alpha_{x}$, $\alpha_{y}$)
and horizontal and vertical $\beta$ functions at the IP ($\beta_{x}$, 
$\beta_{y}$). 
The other five knobs are built using vertical sextupole displacements, 
and they control vertical dispersion ($<y,\delta$E$>$), vertical 
angular dispersion ($<y',\delta$E$>$), and couplings 
($<x',y>$, $<x,y>$,$<x',y'>$). 
\begin{figure}
\centerline{\includegraphics[width=0.8\columnwidth]
{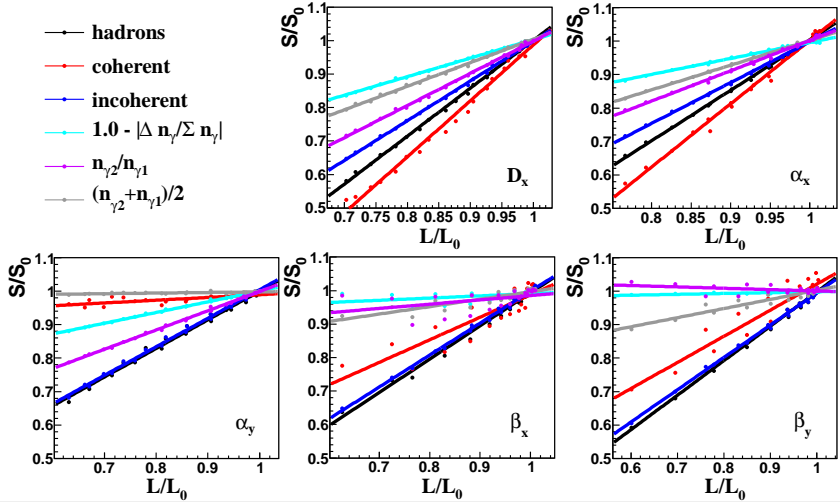}}
\caption{Beam-beam backgrounds signals correlation with total luminosity
  for the scan of the five horizontal knobs.}\label{Fig:2}
\end{figure}

\begin{wrapfigure}{r}{0.5\columnwidth}
\centerline{\includegraphics[width=0.5\columnwidth]
{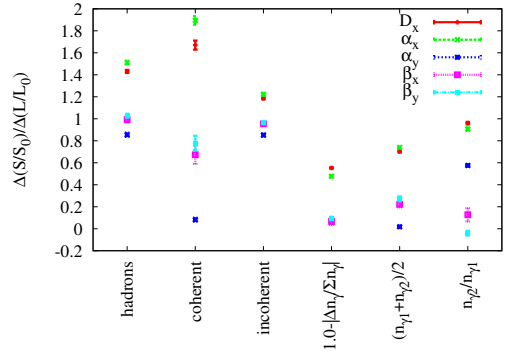}}
\caption{Relative change of the six signals with total luminosity coming
  from the angular coefficient of the fit shown in Fig.~\ref{Fig:2}.}
\label{Fig:3}
\end{wrapfigure}

Figure~\ref{Fig:2} shows a fairly linear correlation of the six signals 
to the total luminosity in this range of scan of the knobs.
The angular coefficients ($\Delta$(S/S$_{0}$)/$\Delta$(L/L$_{0}$)) 
obtained from the linear fits of the data points are shown in 
Fig.~\ref{Fig:3}.
The incoherent pairs signal presents an angular coefficient close to 
one (within 20$\%$ uncertainty) for all the five horizontal aberrations
considered.
The relative hadronic events rate shows the same behavior as the 
incoherent pairs except for horizontal dispersion and horizontal 
waist-shift.   
In these cases the number of $\gamma\gamma$ collisions reduce faster
than the incoherent pairs due to the larger reduction of average 
beamstrahlung photons (i.e. number of $\gamma$ available for collisions).
The relative change of these two types of processes follow the relative
luminosity change independently of the beam aberrations that causes the 
luminosity loss within the 20$\%$ uncertainty. Therefore, incoherent 
pairs and hadronic events can provide an absolute luminosity measurement. 
With absolute luminosity measurement we mean here a signal whose rate 
changes proportionally to the luminosity, regardless of the 
aberrations considered.  
The correlation of the coherent processes with luminosity instead 
assumes quite different values according to the knob (i.e. aberration)
considered.  Therefore, signals from beamstrahlung photons and coherent 
processes, in combination with an absolute luminosity measurement, can 
be used to identify the main aberration of the two beams at the 
collision point, in dedicated feedback. 
The scan of the 5 vertical knobs gives similar results.
 
In practice, it is critical to define a signal that can be  
easily identified against the other processes.
Experimental techniques to detect beamstrahlung photons in the CLIC 
post collision line can be found in~\cite{Rob11}. 
The incoherent pairs are produced with relative small angles with 
respect to the beam axis, but are deflected by the beam fields. 
Therefore, the pairs particles can have large angles. The integration 
of pairs energy above a certain angle with respect to the beam axis 
has been studied as potential signal for luminosity optimization in 
~\cite{Dani04}. In CLIC their detection could be more complicated 
due to the presence of the coherent pairs in the forward 
region, leptons coming from hadronic events and Bhabhas.
In the following we discuss further the possibility to define
a trigger using hadronic events by looking in particular at its
multiplicity in the final state of the process.

\section{Hadronic Events}

Hadrons at linear colliders are produced by the process 
$e^{+}e^{-}$ $\to$ $\gamma \gamma$ $\to$ hadrons.
The total $\gamma \gamma$ $\to$ hadrons cross section is known 
experimentally up to 200 GeV in the center of mass energy. The simplest
model of the energy-dependence of the $\gamma \gamma$ $\to$ hadrons 
cross section ($\sigma$) is the vector meson dominance one. The model 
assumes that the photon resonates to a hadronic state (a $\rho$) with a
certain probability~\cite{PRD94}, with the energy dependence expressed
as:

\begin{equation}
  \sigma = 211\textrm{nb} \cdot \big(\frac{s}{GeV}\big)^{\epsilon} + 215\textrm{nb} \cdot 
  \big(\frac{s}{GeV}\big)^{\mu} 
\end{equation}

where $\epsilon=0.0808$ and $\mu=0.4525$~\cite{Sjoe96}.
GUINEA-PIG implements the above parametrization of the total 
$\gamma \gamma$ $\to$ hadrons cross section. An electron or positron is 
replaced by the appropriate number of photons from the equivalent 
spectrum. 
The energies of the two colliding photons can be stored in a file and then
loaded into PYTHIA~\cite{PYTH}, or an equivalent code, to generate the hadrons. 
 
In order to define a region where the hadrons multiplicity can be detected 
two different $p_{T}$ cuts are applied to the 
charged particles. This ensure that they can travel in the 
forward detector region or in the detector main tracking region, 
considering a B-field of 5 Tesla. Following~\cite{detacc} we 
consider tracks with $p_{T} > $ 0.050 GeV and 27 mrad $< \theta <$ 117 
mrad for the forward region and tracks with $p_{T} >$ 0.160 GeV and 117
mrad $< \theta <$ 1.57 rad for the main tracking region.
The 27 mrad condition for the forward region 
is due to the envelope of the incoherent pairs while traveling in 
the detector solenoid magnetic field. The aim is to define a  
``background free'' region to improve the quality of the identification of
the signal against the other background sources.
\begin{figure}
  \begin{minipage}[b]{7.1cm}
    \includegraphics[clip,height=7.cm]{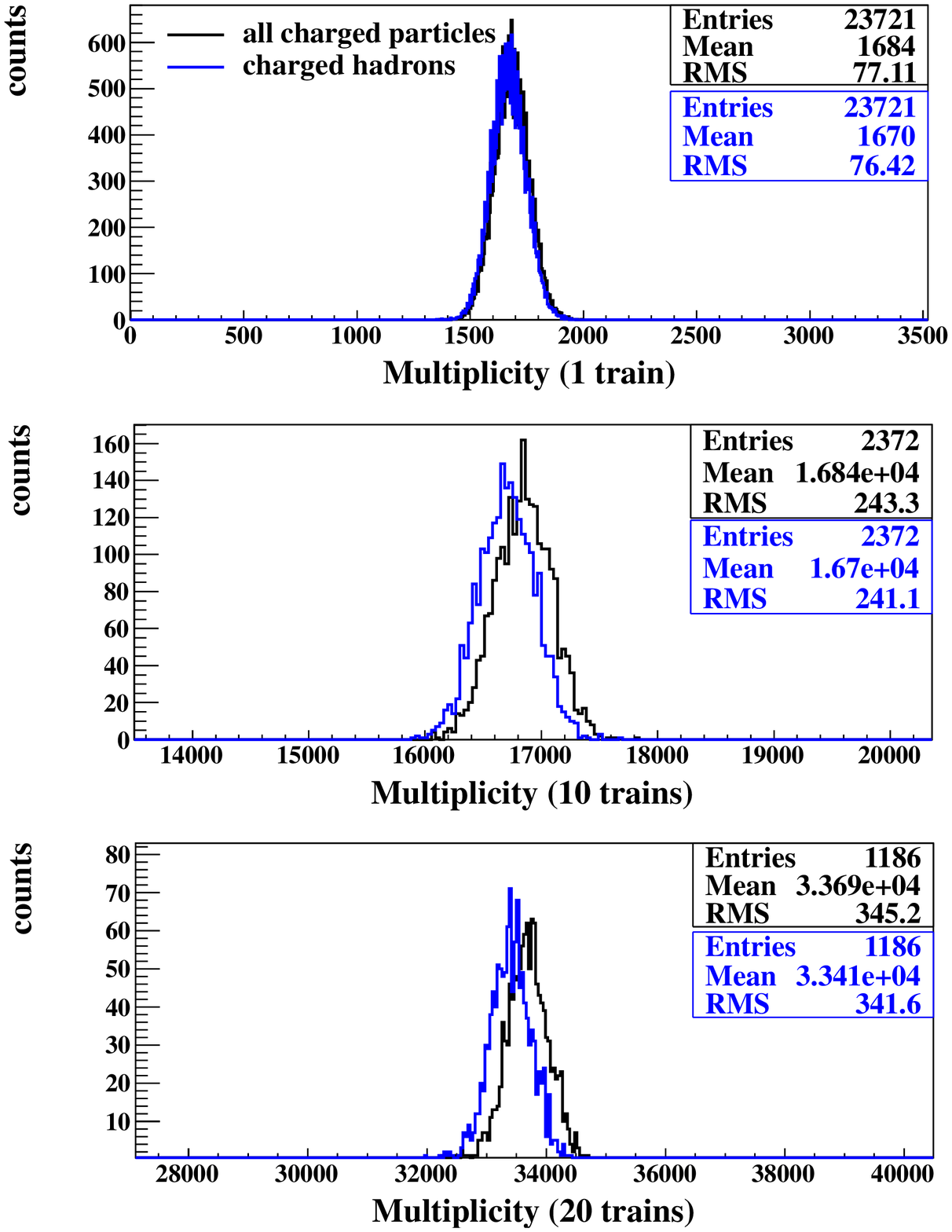}
  \end{minipage}
  \begin{minipage}[b]{7.1cm}
    \includegraphics[clip,height=7.cm]{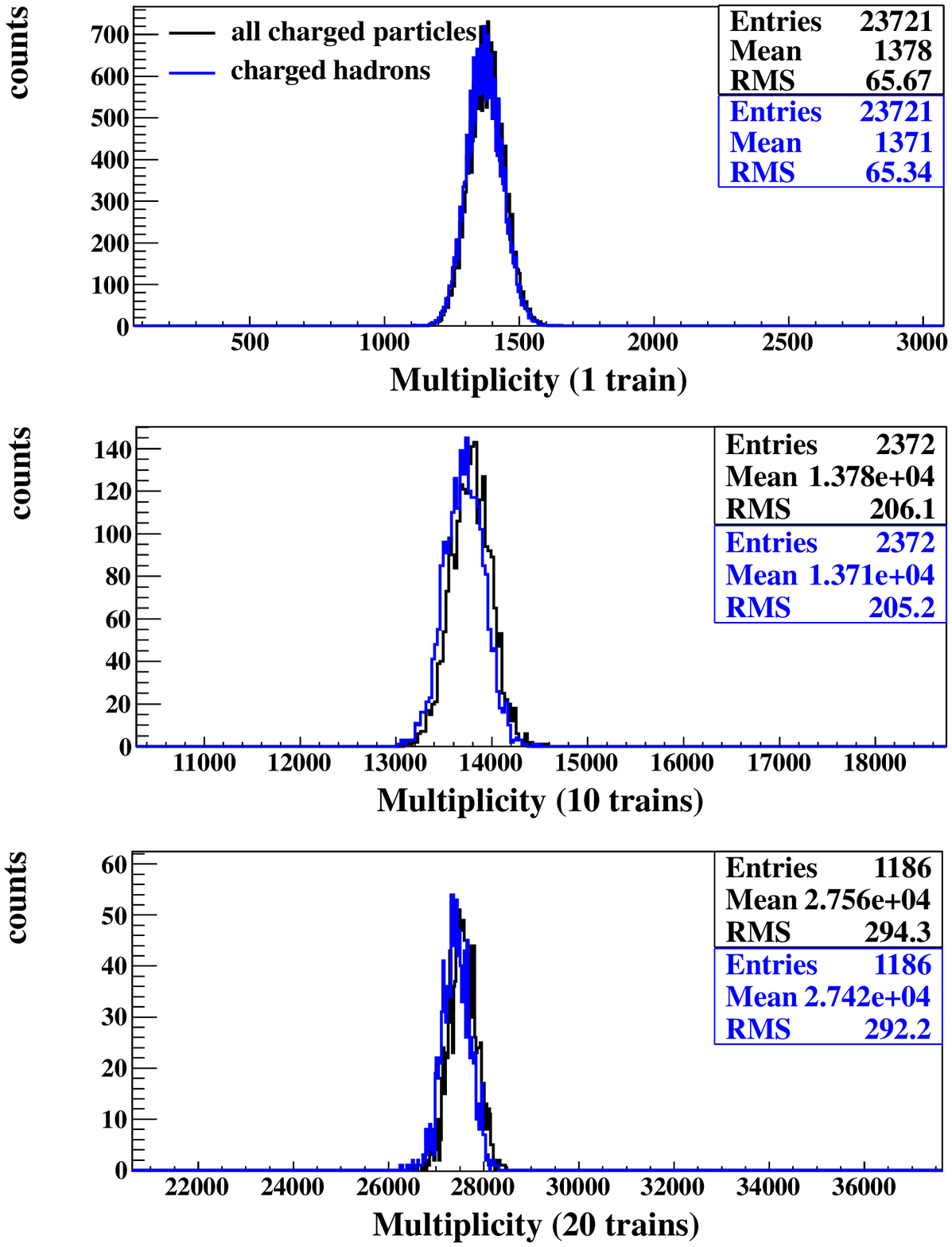}
  \end{minipage}
  \caption{ Integrated charged particles and charged hadrons multiplicity
    over 1,10 and 20 CLIC trains with $p_{T} >$ 0.050 GeV/c and 0.027 
    $< \theta <$ 0.117 rad (left), and with $p_T >$ 0.160 GeV/c and 0.117 
    $< \theta <$ 1.57 rad (right).}
  \label{Fig:5}
\end{figure}
The resulting multiplicity distributions of all the charged particles 
and that of the hadrons, according to the selected angles and momenta,
are shown in Fig.~\ref{Fig:5}.
The multiplicities are integrated over 1, 10 and 20 trains. 
Almost all the multiplicity from $\gamma \gamma$ collision consists of
charged hadrons. The detection of the hadronic component can be 
optimized in order to identify them against the lepton component or 
leptons coming from other processes.
The mean value of the distribution over 20 trains is determined with 
about 1$\%$ fluctuation, which gives about 1.2$\%$ precision in the 
absolute luminosity measurement (due to the 20$\%$ uncertainty we 
discussed before). 
Given the CLIC repetition rate (50 Hz) 20 trains correspond to 0.4 
seconds.
Requiring a total time of $\le$ 0.1 s for the read-out electronics and
signal elaboration, one luminosity measurement should take about 0.5 s.
Taking into account the number of luminosity
measurement needed the total time to tune the BDS, starting from
10 $\mu$m pre-alignment of the magnets, would be 
about 15 min when the BBA technique in combination with the FFS 
sextupoles knobs are applied, even if with low success rate. The total
time required to tune the BDS with the luminosity optimization 
technique is instead of the order of 2 hours.
Note that one full scan of the ten knobs takes
$\sim$ 3 min, which is compatible with
the requirement for the mitigation of the dynamic imperfections.  
The full CLIC detector model is not considered in these simulations.
The actual amount of material and the interaction of these particles
with matter should be considered in order to define the best region 
of detection of the signal, and minimize the number of bunches to sum
in the trigger.

\section{Conclusions}

In order to mitigate the impact of magnet displacements in the CLIC BDS
different techniques are compared. The best results reached so far
are obtained by combining luminosity optimization and sextupole
knobs scan: 90$\%$ of the considered machines reach 90$\%$ of CLIC 
nominal luminosity. 
In particular tuning knobs exploiting the sextupoles of the FFS have 
been proven a powerful tool to recover the luminosity loss due to 
magnet displacements.
Tuning knobs and the luminosity optimization technique require a fast 
luminosity measurement. 
For this purpose the possibility to use $ \gamma \gamma \to$ hadrons 
background is investigated. The charged particles multiplicity 
from this process in the vertex-tracking and/or in the forward region 
of the detector could provide a signal for a fast luminosity 
measurement in less than 1 s with $\sim$ 1$\%$ precision. 
Given the number of luminosity measurements needed by the different 
alignment techniques here considered, the full tuning of the CLIC BDS
against magnet displacements can be achieved in the range between about
15 min and 2 hours.  

\begin{footnotesize}

\end{footnotesize}
\end{document}